%% file: main.tex
\documentclass[conference]{IEEEtran}
\IEEEoverridecommandlockouts
\usepackage{cite}
\usepackage{amsmath,amssymb,amsfonts}
\usepackage{svg}
\usepackage{algorithmic}
\usepackage{graphicx}
\usepackage{textcomp}
\usepackage{xcolor}
\usepackage{todonotes}
\usepackage{url}
\newcommand{\lmttfont}{\fontfamily{lmtt}\selectfont}
\def\BibTeX{{\rm B\kern-.05em{\sc i\kern-.025em b}\kern-.08em
    T\kern-.1667em\lower.7ex\hbox{E}\kern-.125emX}}
\begin{document}

\title{Security Assessment and Hardening of \\ Fog Computing Systems}

\author{\IEEEauthorblockN{Carmine Cesarano}
\IEEEauthorblockA{\textit{Università degli Studi di Napoli Federico II, Italy} \\
carmine.cesarano2@unina.it}

}

\maketitle

\begin{abstract}
In recent years, there has been a shift in computing architectures, moving away from centralized cloud computing towards decentralized edge and fog computing. This shift is driven by factors such as the increasing volume of data generated at the edge, the growing demand for real-time processing and low-latency applications, and the need for improved privacy and data locality. Although this new paradigm offers numerous advantages, it also introduces significant security and reliability challenges. This paper aims to review the architectures and technologies employed in fog computing and identify opportunities for developing novel security assessment and security hardening techniques. These techniques include secure configuration and debloating to enhance the security of middleware, testing techniques to assess secure communication mechanisms, and automated rehosting to speed up the security testing of embedded firmware.
\end{abstract}


\section{Introduction}
\label{sec:introduction}
\input{1_introduction}

\section{Background}
\label{sec:backgroun}
\input{2_background}

\section{Proposal}
\label{sec:proposal}
\input{3_proposal}

\section{Conclusion}
\label{sec:conclusion}
\input{4_conclusion}

\section*{Acknowledgement}
Professor Roberto Natella supervises the dissertation research. This work has been partially supported by the MUR PRIN 2022 program, project FLEGREA (CUP E53D23007950001) and by the PhD Program in Information Technology and Electrical Engineering (ITEE) of the Federico II University.

\bibliography{bibliography} 
\bibliographystyle{ieeetr}

\vspace{12pt}

\end{document}

%% file: 1_introduction.tex
% Paradigm Shif: Limitazioni del vecchio, vantaggi del nuovo
In recent years, the computing landscape has experienced a notable transformation, with the emergence of Edge and Fog Computing as alternatives to traditional Cloud Computing. This shift is driven by the limitations of the cloud paradigm, which, despite advancements in performance, speed, and availability, struggles to meet the increasing demands placed upon it. The need for real-time data processing, low-latency applications, and optimal resource utilization has fueled the widespread adoption of Edge and Fog Computing.

% Fog, Edge computing: caratteristiche e contesti applicativi
In the traditional Cloud Computing paradigm, IoT devices serve as the data source and are completely decoupled from the computing layer, which relies on cloud resources.
However, the emerging Edge and Fog paradigms provide a compelling alternative by decentralizing the computing layer and bringing storage, networking, computing, and data management capabilities closer to the data source. This proximity brings numerous benefits, such as reduced latencies, lower bandwidth requirements, and improved security. By processing large volumes of data at the network edge, these paradigms enable real-time decision-making, support resource-constrained devices, and enhance overall system efficiency. Consequently, these innovative paradigms have found widespread application for Cyber-Physical Systems, IoT, and Industrial IoT (IIoT) in critical domains. For instance, in areas like smart cities \cite{tang2017incorporating}, vehicular systems \cite{du2020new}, industrial automation \cite{pop2018enabling} and healthcare \cite{paul2018fog}, fog computing represents a solution to improve performance and security.

% Security issues introdotti dallo shift
However, this shift towards edge and fog computing introduces new security challenges that must be addressed. With the decentralization of resources, the attack surface expands, making these distributed systems vulnerable to various threats. The dynamic and heterogeneous nature of edge and fog computing environments further complicates security measures, as they often rely on diverse hardware platforms, operating systems, and middleware. Malicious actors can exploit vulnerabilities in edge devices, communication channels, and data storage to compromise the system's integrity, confidentiality, and availability.

% Esempi di attacchi reali
Recent attacks like Mirai and IoT Reaper have shown the urgent need to bolster the security posture of edge and fog computing architecture. Mirai, a notorious malware that emerged in 2016, targeted vulnerable IoT devices by exploiting weak or default credentials. By infecting a vast network of compromised devices, Mirai orchestrated distributed denial-of-service (DDoS) attacks, disrupting online services, and impacting the reliability and availability of critical infrastructure. Similarly, IoT Reaper, in 2017 exploited vulnerabilities in IoT devices to create a botnet capable of launching various types of attacks.

% Necessità di tecniche di hardening e assessment
Given the critical role of edge and fog computing in various operational contexts, it becomes paramount to design efficient hardening and assessment techniques for these architectures. Securing and isolating environments at multiple, distributed, and multi-tenant nodes is crucial to mitigate vulnerabilities. Moreover, safeguarding interactions between components helps protect against attacks and unauthorized access. 

% Introduzione dei contributi
This paper reviews the architecture and technology stack for the fog computing paradigm and related threats. Additionally, it proposes a comprehensive plan for security hardening and assessment of these architectures. The plan aims to different techniques in the following steps:
\begin{itemize}
    \item Security hardening of middleware, including secure configuration and debloating.
    \item Security testing for secure communication mechanisms.
    \item Automated rehosting to speed up firmware security \\ testing.
\end{itemize}

By following these steps, the dissertation research will provide techniques and methodologies to enhance the security posture of edge and fog computing architecture. This plays a critical role in fostering the adoption and advancement of this new computing paradigm in various operational contexts.

%% file: 2_background.tex
\begin{figure*}[h]
    %\vspace{-10pt}
    \centerline{\includegraphics[width=2\columnwidth]{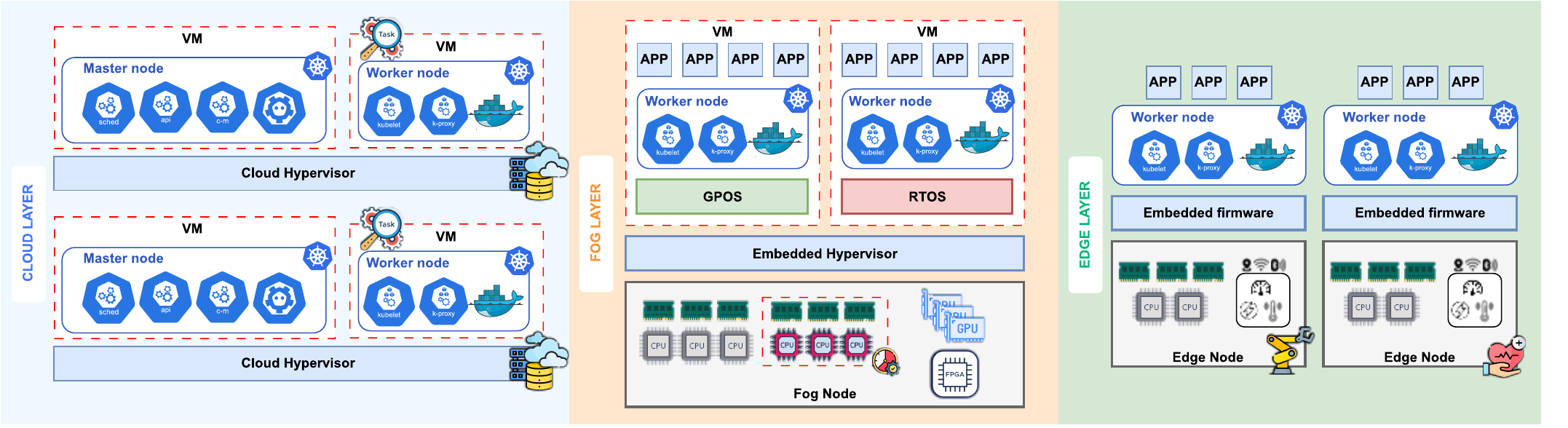}}
    \caption{Example of a Fog Computing architecture 
    \label{fig:architecture}}
\end{figure*}

Compared to traditional cloud infrastructures, edge and fog computing architectures have unique elements and capabilities. 

\bigbreak
% Edge computing (archittura, caratteristiche, security issues)
\noindent \textbf{Edge computing} architectures consist of two layers: the cloud layer and the edge layer. The edge layer comprises edge nodes including industrial sensors, computing machinery equipment, automotive control systems, consumer electronics, and wearables. These nodes have small size, weight, power, and cost (SWaP-C) factors, and are equipped with microprocessors, microcontrollers, Digital Signal Processors, System-on-Chip technology, and input/output and memory resources. The edge layer enables data processing to be performed in close proximity to the data source. Filtering and clustering at the edge minimize the need to transfer massive amounts of data, which can also be security-sensitive, to the cloud.

% Fog computing (archittura, caratteristiche, security issues)
\bigbreak
\noindent \textbf{Fog computing} extends the concept of edge computing by introducing an intermediate fog layer between the edge layer and the cloud layer. Fog nodes, which can be edge servers, are equipped with specialized or general-purpose hardware, such as CPUs, GPUs, real-time processors, and FPGAs. These fog nodes prioritize computational power over SWaP factors. They handle resource-intensive tasks such as data collection from edge nodes, aggregation, preprocessing operations, data analytics, and caching. The fog layer adds computing capabilities along the path between the data source and the cloud, as well as enabling offloading of resource-intensive tasks to the cloud.

% Security threats
\bigbreak
\noindent \textbf{Security threats} are introduced by the adoption of fog and edge computing, due to the unique characteristics of these architectures:
\begin{itemize}
    \item \textit{Threats to edge and fog nodes}: the distributed environment and deployment at the network edge increase endpoints, expanding the attack surface. Attackers target these endpoints for unauthorized access, launch attacks, and take advantage of security weaknesses against other tenants in the infrastructure. Limited resources hinder the implementation of robust security mechanisms, potentially leading to security breaches. The use of open-source software and rapid development cycles may result in deploying devices with vulnerabilities. Therefore, it is important to have an integrated secure supply chain, updating, and testing processes in place.    
    \item \textit{Threats to communication channels}: with geographically distributed entities and interactions, there is an increased risk of communication interception, tampering, and data breaches. Ensuring the security of communication between fog and edge devices becomes crucial to prevent unauthorized access and maintain data integrity.
\end{itemize}

Overall, the characteristics of fog and edge computing architectures introduce security challenges that need to be addressed to protect against various threats and maintain the security and integrity of the system.

\bigbreak
\noindent \textbf{Virtualization and orchestration} technologies play a crucial role in fog and edge computing systems, ensuring isolation and efficient management. They support essential characteristics such as multi-tenancy, high reliability, distributed deployment, performance optimization, and dynamic application workload. Fig. ~\ref{fig:architecture} shows a three-tier architecture implemented on a cluster of heterogeneous edge and fog nodes and cloud resources. In this architecture embedded hypervisors like ACRN, Bao, or Jailhouse are used to consolidate multiple isolated virtual machines (VMs) onto a single physical fog node. Within each virtual machine, containerization provides further isolation, encapsulating applications and their dependencies. Cloud hypervisors such as VMware vSphere, Xen, and KVM, on the other hand, provide VMs suitable for resource-intensive tasks in the cloud layer.

In addition, the cloud layer can provide orchestration capabilities for the fog and edge nodes. As shown in Fig. ~\ref{fig:architecture}, the architecture includes some cloud-based VMs hosting the \textit{Kubernetes master nodes} for the orchestration control plane. Kubernetes is a middleware system that automates various tasks related to application lifecycle management. These tasks include deployment, provisioning, scheduling, scaling, and load balancing of containerized applications or services. By efficiently distributing and managing containers, the orchestrator ensures the desired state while considering the application workload and availability of system resources. In the reference architecture, the orchestrator can utilize both physical and virtual resources as \textit{Kubernetes worker nodes} for running containers. Physical nodes consist of embedded devices within the edge layer, while fog servers can employ hypervisors to provide multiple virtual nodes.

%% file: 3_proposal.tex
The research activity presented in this paper aims to develop security hardening and assessment techniques for the building blocks of the fog computing architecture described in Section \ref{sec:backgroun}. The development plan consists of the following steps:

\begin{itemize}
    \item Developing methodologies to enhance the security of open-source middleware, including secure configurations and reducing the attack surface through debloating. This step focuses on strengthening isolation properties to prevent security breaches.
    \item Developing security testing techniques to assess communication mechanisms. This step focuses on enforcing mechanisms and policies for secure communication between isolated environments. 
    \item Developing techniques to automate the rehosting of embedded firmware. This step focuses on speeding up firmware testing to identify vulnerabilities.
\end{itemize}

\subsection{Secure configuration and debloating}
\label{sec:debloating}
Edge and Fog Computing architectures employ complex and highly configurable middleware systems, comprising multiple components. Orchestrators, mentioned in Section \ref{sec:backgroun}, serve as an example of such middleware systems. Configurations of specific components can enable or disable features, which can have an impact on the attack surface. Employing secure configuration and software debloating techniques can enhance the security posture of the system and effectively reduce the attack surface. 
In this phase of the research activity, the focus will be on developing efficient methodologies to explore the configuration space of orchestrators like Kubernetes. The goal is to identify security-specific configurations, including disabling unnecessary services, configuring access controls, employing strong encryption, enabling firewalls, and implementing secure authentication mechanisms. Natural language processing can assist in this task by classifying configurations and identifying security-related ones. The knowledge base for this research includes configuration definitions and descriptions in the code. Additionally, information can be retrieved from versioning systems, analyzing reports and comments associated with security-related commits that were released to fix known vulnerabilities. Best practices, guidelines, and security recommendations provided by vendors, security organizations, and industry standards (such as the CIS Benchmark \cite{CISbenchmark} in the case of Kubernetes) can also be used.
The goal of a secure configuration is to minimize the attack surface, but disabled components or features can still be present in the software, especially with runtime configurations. To enhance security, debloating techniques have been proven effective in eliminating unnecessary or potentially vulnerable components \cite{brown2019less}. Various lightweight debloated distributions exist for systems like Kubernetes, including k0s \cite{k0sproject}, k3s \cite{k3sproject}, microK8s \cite{microk8s}. However, it is worth noting that these distributions primarily aim to enhance performance in resource-constrained environments, rather than being explicitly security-oriented. In this step, a configuration-driven debloating approach will be used to identify and remove functionalities or components that are known to never be executed. For example, in Kubernetes, it is possible to configure different types of volume. Using static taint analysis, we can identify and debloat all volume drivers except the one configured, based on the configured options provided.

\subsection{Inter-environment secure communication}
\label{sec:communication}
Hypervisors and Containers provide different levels of isolation to ensure secure and multi-tenant utilization of fog computing architecture. However, in distributed contexts, there are scenarios where applications require relaxed isolation to enable data sharing, API invocations, or distributed processing. In fog computing, isolated environments can communicate with each other using various mechanisms provided by operating systems, firmware, middleware, and hypervisors. Networking capabilities enable communication over local networks or the Internet. Inter-process communication (IPC) facilitates data exchange and API invocation through sockets or shared memory. Service mesh technologies, such as Istio \cite{istio} or Linkerd \cite{linkerd}, can enhance communication by managing interactions between services.

Implementing security mechanisms and policies is crucial to ensure secure communication and protect the resources involved. Firewalls play a significant role in filtering network traffic via application-level policies. Access Control Lists (ACLs) are used to define permissions to restrict communication between different privilege levels. Additionally, SELinux enforces fine-grained restrictions on communication by implementing mandatory access control (MAC) policies.

During this phase of the research activity, the objective is to assess the communication system and identify vulnerabilities in the employed security mechanisms. One approach is to develop techniques that virtualize the system under test and utilize the introspection capabilities of a hypervisor. By using introspection, it is possible to observe the behavior of the entire system and the interactions between isolated environments. The integration of fuzzing techniques and the TCG plugins of the QEMU Emulator can be utilized in this approach. This allows for the interception and transparent corruption of exchanged messages, with the aim of discovering potential vulnerabilities in the secure communication mechanisms.

\subsection{Firmware rehosting}
\label{sec:rehosting}
Embedded firmware in the edge nodes of the fog computing architecture can be vulnerable to various software attacks, which makes firmware testing a crucial aspect in ensuring the security and reliability of the entire software stack. 

\textbf{Rehosting techniques} provide a solution by emulating the complete hardware system, including CPUs and peripherals, enabling dynamic testing methods like fuzzing and symbolic execution. However, the high interaction between firmware and hardware peripherals poses challenges for effective testing, due to the heterogeneity of devices and peripherals. Hardware-in-the-loop solutions, proposed by Koscher et al. \cite{koscher2015surrogates}, Zaddach et al. \cite{zaddach2014avatar}, and Kammerstetter et al. \cite{kammerstetter2014prospect}, aim to address this challenge by forwarding interactions to real hardware. However, this approach requires expensive hardware and carries a risk of physical system damage. Gustafson et al. \cite{gustafson2019toward} and Spensky et al. \cite{spensky2021conware} have applied machine-learning techniques for automated peripheral emulation, reducing the dependency on real hardware to only the training phase. However, current solutions mainly focus on modeling peripherals with simple behaviors and firmware interactions. Overcoming this limitation is crucial to effectively rehost complex peripheral architectures and simplifying security testing for embedded firmware.

This research phase aims to develop techniques for accurately emulating complex peripherals using a machine learning-based code generator. During an initial \textit{learning phase}, representative workloads are used to stimulate the firmware running on the real hardware. The behavior of already emulated hardware peripherals is profiled and output logs are recorded. In this phase, the micro-operations of the peripheral are linked to corresponding software implementations. This phase builds a dataset composed of a set of samples {\lmttfont <micro-op pattern, micro-op implementations>} that represents the profiled peripheral behavior. In the \textit{inference phase}, the dataset is used to train a code generator for new unemulated peripherals. The output logs are profiled during the firmware execution on the real hardware, micro-operation models are generated, and the code generator translates them into software implementations. State-of-the-art emulators usually cover all peripheral classes and peripherals from the same classes share micro-operations, making the technique applicable to a significant portion of non-emulated peripherals. This automated approach simplifies security testing for embedded firmware by accurately emulating complex peripherals.

\subsection{Status of the dissertation research}
The first and second years of the research activity will primarily focus on developing debloating techniques for open-source middleware and assessing mechanisms for secure communication. Preliminary analyses were conducted to evaluate the feasibility of a configuration-drive debloating technique, specifically targeting Kubernetes. Furthermore, a fuzzing testing technique was designed and implemented, leveraging the QEMU emulator, to test inter-process communication in the Multiple Level of Security (MILS) architecture. The third year of the research activity will be dedicated to the development of an automated modeling technique for embedded device peripherals.

%% file: 4_conclusion.tex
This research paper proposes a plan to enhance the security of fog computing systems. By focusing on secure configuration, inter-environment secure communication, and firmware rehosting, the proposed methodologies aim to strengthen the overall security posture of these complex systems. Research activities have demonstrated promising results in developing debloating techniques for open-source middleware and assessing secure communication mechanisms using fuzzing testing. Future work will focus on advancing automatic modeling techniques for embedded device peripherals. This research contributes significantly to the growth of knowledge surrounding the security of fog computing, and the proposed methodologies have the potential to greatly benefit the industry as a whole.